# A Count of Palm Trees from Satellite Image

[1] Emad Ali Al-helaly, [2] Noor Ali Al-Helaly

[1] Kufa University, College of Engineerin
Email: imada.alhilali@uokufa.edu.iq ; EmaadAlhilaly@gmail.com
[2] Al-Emam Al-Kadhem University, College of Education
Email: nnnaaa90@gmail.com

**Abstract.** In this research the number of palm trees was calculated from the satellite image programmatically, taking advantage of the accuracy of the spatial resolution of satellite image, the abilities of software recognition, and characteristics of the palm tree, which give it a systematic top view can be distinguished from the satellite image and the manner of cultivation and vertical growth and stability form for long periods of time. While other trees are irregular in shape mostly because of their twisted branches. Palm trees consist of a long stem, a large head, and a large flare that is almost circular and consists of large tufts. The palms have large self-shadows other than ordinary leaves. The large shadows and the circular shape of the upper view give it a special feature that we could use to design a program that distinguishes the shape of the palm without all the trees. Then it counts the number of palms in any field shown in the satellite image. This method is useful in counting the number of palm trees for commercial, agricultural or environmental purposes. It is also can be applied to high resolution satellite imagery such as QuickBird because the resolution of the images is 0.6 meters. Less accurate images such as the 10-meter SPOT do not show the interior shadows of the top view of the palm enough, nor the accurate satellites (5 meters), while the interior shadows appear in high-resolution images only (0.6 meters) or below. It can also be applied to aerial images of less capacity because they are more accurate of course. Satellite images can be obtained free from Google Earth explorer, which can be downloaded free from the Google web site. It connects the user to a global database of high resolution images for all regions of the world.

## 1 Introduction

Palm trees are widespread in southern Iraq and in a number of countries with warm climates and abundant water. They are fruit trees, producing wet [R.1] , a useful and important food

[R.2] , jam, molasses and others. Palm trees can also be utilized from their palms, kebabs and stems for other purposes [R.3]. It is considered one of the natural resources in some countries, as well as it is one of the productive and expensive crops, Palm groves are sold according to their size and the date palm trees they contain. The number of palms is usually calculated in a traditional way based on the individual tree account, a tedious method in large areas, and in the wide statistics of government departments and research whose date numbers are calculated within the study areas. Here we have introduced a software method to calculate the number of palm images using modern software and proper algorithms to distinguish the top view shape of the palm, and can be applied to this method with high accuracy and speed and digital expressions can be applied to other statistics based on the number of palms.

## 2    The Cast Shadow and Self Shadow of Objects

The light travels in straight lines and, if faced by the dark objects, it leaves behind an external shadow called cast shadow. It is seen in the satellite image as a separate landmark from the original shaded body. There is an inner shadow called the self-shadow [R.4]. The self-shadow - in terms of shape and opacity - results from the nature of the darkened surface of the body and the angle of the fall of light rays on it. The smooth surface has gradient self-shadows from the illuminated area to the dark, the surface with sharp terrain has a sharp shadow, and the surface with roughness or roughness has light shadow that may not be noticeable other than if the topography is prominent and large, they make sharp and clear self-shadows. The self-shadow is a distinguishing feature of the shape and its topography. The more precise the shadows are, the more details can be distinguished from the others, especially in objects that are depicted from one angle.

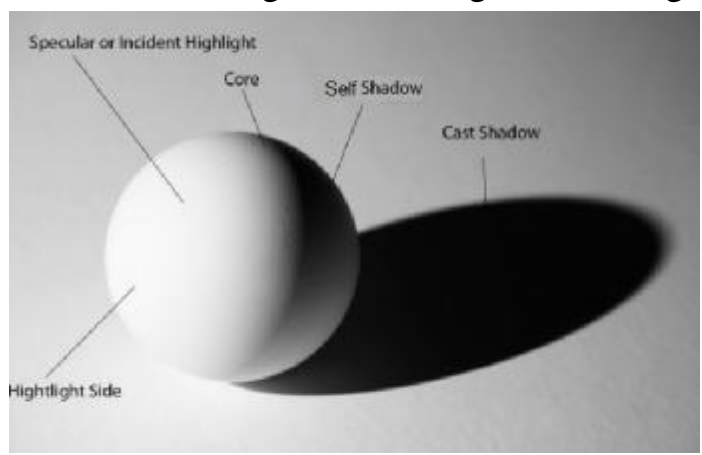

**Figure 1 The Cast Shadow and Self Shadow**

## 3      Image Resolution

The satellite image is the images taken from satellites by high-resolution sensors on the satellites which orbit the space at a height of hundreds of kilometers, the satellite image taken covers vast areas estimated at hundreds of square kilometers due to the high altitude of the satellite [R.5]. The high latitude of the satellite affects the accuracy of images, this is noted since the first space images of the 1970s, $30 \times 30$ meters, such as the Landsat satellites, which can distinguish objects with a width greater than 30 meters. In the 1980s, the SOPT sensors of (20 meters) or (10 meters) were launched, but in the 1990s the accuracy of the images resolution increased to modern electronic and optical techniques until the accuracy of the satellite images to (0.6 meters) [R.5].

## 4      Recognition of Shadows from Satellite Images
Low resolution satellite images (80, 30, ... m) or less show only the large shadows, such as mountains, clouds and large buildings. The self-shadows may disappear, and only the large objects appear in the picture. High resolution images (0.6 meters, 0.41) or more show cast shadows and some large self-shadows larger than 60 centimeters in size.

## 5 The Self Shadow of Palm Head

Palm trees are large in size, and have external shadows, but the outer shadows do not benefit us in distinguishing them because they disappear among the other trees, but their self shadows are created from the shadows of the palm rings. The ring consists of a large number of small feathers that branch out almost systematically from both sides of the central stick that are connected to the palm head, and the length of the ring is usually several meters. The dimming of the rest of the palms can be distinguished from high-resolution satellite images [R.6]. The shape of the palm head from the top view is almost circular and the distribution of the palm from all sides is systematic as well in the grove, allowing for a regularity that can be distinguished by the techniques of discrimination.

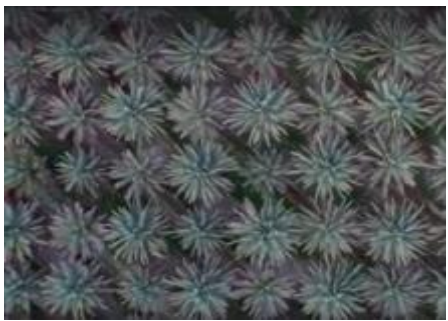 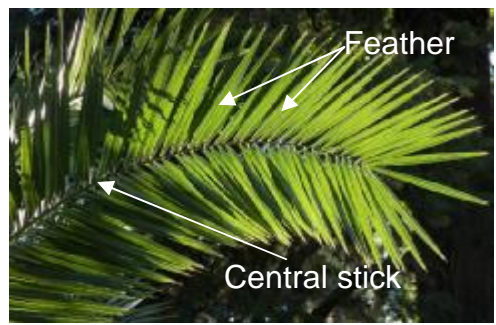

**Figure 2  Palm Grove Upper View.**        **Figure (3) Ringworm of Palm tree.**

Palm trees usually grow vertically and are sporadically grown, allowing palm trees to overlap and skip each other except in exceptional cases and with the growth of the palm separate from any trees that overlap. The palm tree characters allow the shape of the palm to be distinguished from the top view and can take a distinctive imprint that can be observed from the satellite image. These features are not available in most other trees which have small leaves and irregular distribution.

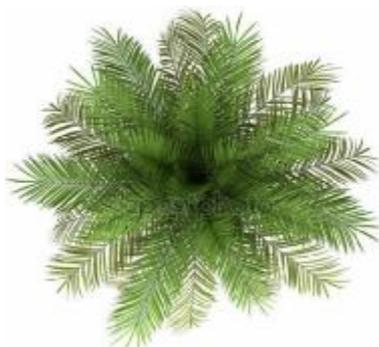 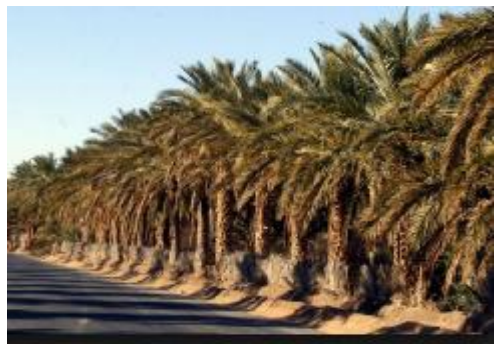

**Figure (4) The Top View of Palm.**        **Figure (5) Side View of Palms.**

# 6     The designed software to recognize palms

The abilities of MATLAB (Release 2013a) were used to design a algorithms to recognize the palm tree depending of the characteristic of pal trees which were submitted in the previous paragraphs [R.7]. The algorithm converts the color image, which includes the overhead image of palm trees, into a gray image and then a binary image to facilitate identification and discrimination tasks. The algorithm then applies a set of morphological processes to fill the gaps, then investigates the specific palm forms mentioned for the purpose of obtaining blocks of connected pixels that often represent palm trees, then removing small elements such as impurities or image residues after converting them into binary images. With the determination of the connected masses that carry the same value, and measuring the properties of the clusters related to center size, borders and others. Then filter the connected blocks depending on the specific properties and especially the space for the purpose of deleting any mass that does not fit the characteristics of the palm. The final step determines only the center and area of each palm tree and is determined by a circular curved figure around it. It senses the number of palms in the picture and can be separated as separate images and stored in a database. In the MATLAB results window, the number of palms appears in the image. The accuracy of this method can be ascertained by applying it to any image while modifying some of the characteristics of the program in the event of specialties of the satellite image. The figure shape below illustrates the procedure of the algorithm.

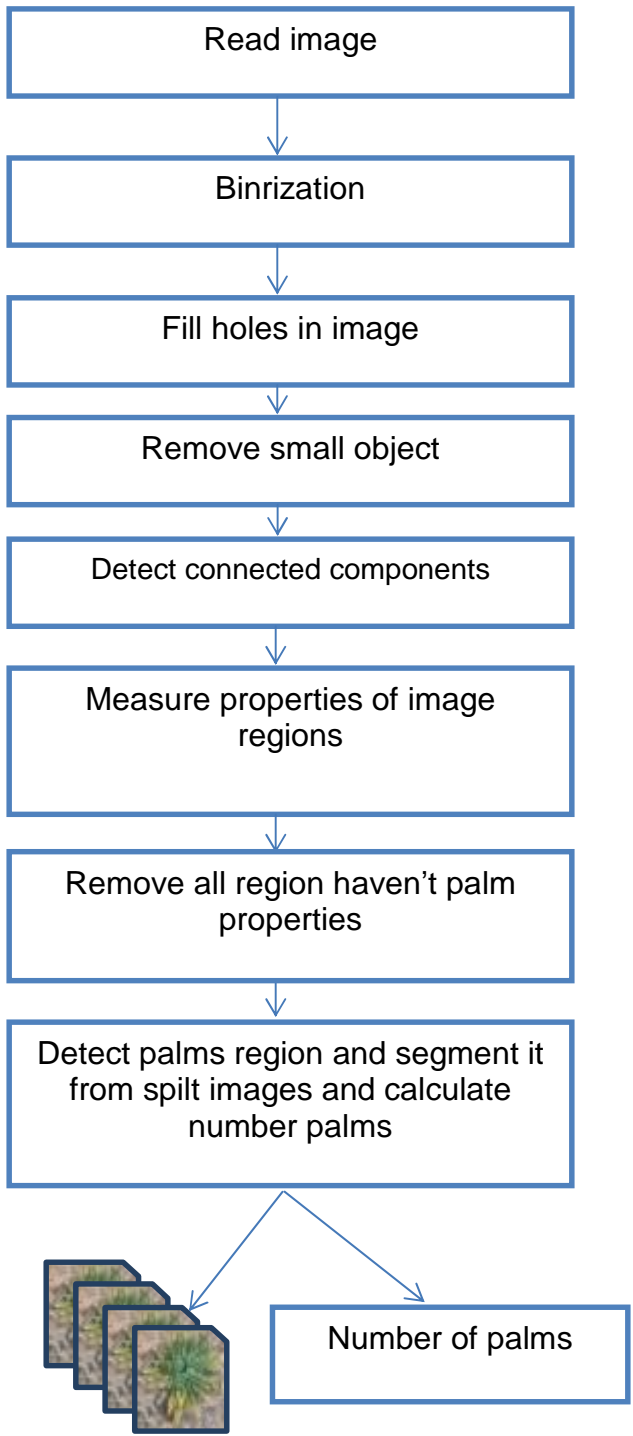

## 7   Applying the Software

The algorithm was applied for testing on simple palm images, and then adopted the distinctive shape of the palm to count the number of palms in the whole image which contain the palm grove. The following is the first application of the program on a palm image from the top.

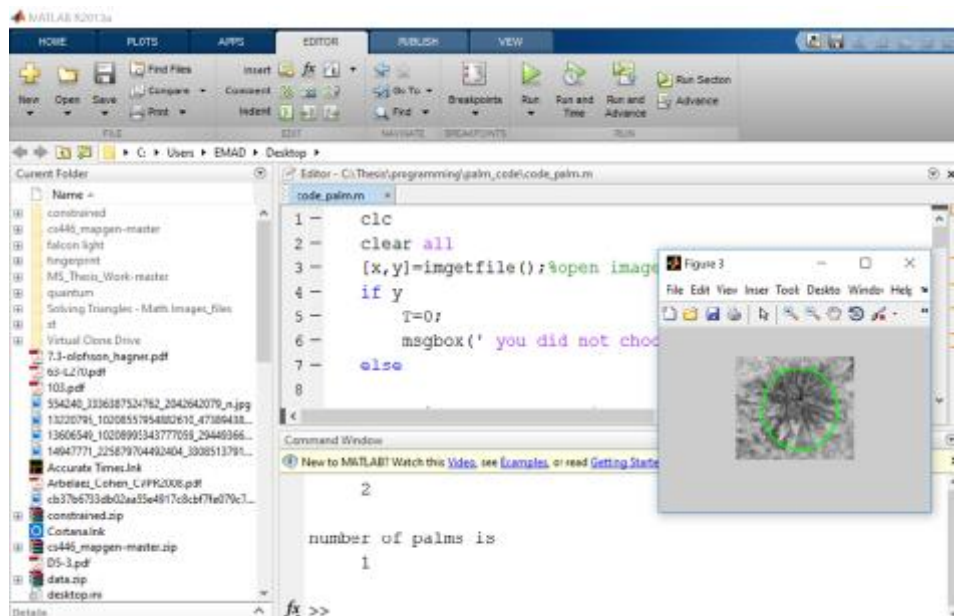

**Figure (8) Applying the Algorithm to Image has Single Palm.**

Figure (9) illustrates the result shown by the algorithm when used in an image that includes a number of palm trees with high resolution QuickBird satellite (0.6 meters). Note in the picture box that the circles around the palm trees represent the sense of the algorithm and its distinguishing feature of the palm. The number in the MATLAB window represents the number of palms in the satellite image.

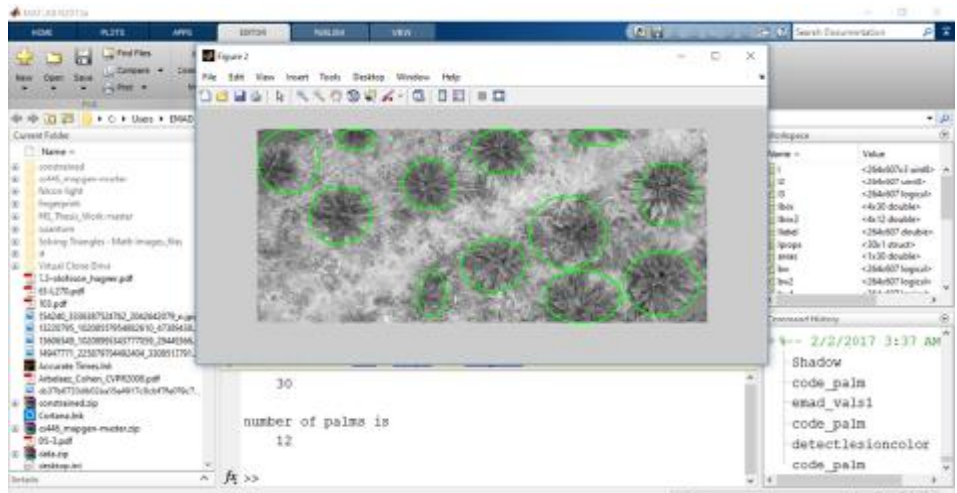
**Figure (9) Applying the algorithm to image has many palms.**

This algorithm allows to calculate the number of palms in any selected area with high accuracy, and we have tested the areas in the above image and reported accurate results verified by calculating the number of palms in the normal account. The algorithm can be more reliable in any other statistics such as estimating the number of dates production by multiplying the number of palm with the average production per palm, or other products, and thus we get accurate statistics at a low cost. Free Google Earth imagery can be produced to produce statistics for any area where the number of palms is to be counted. The same program can be adopted in aerial images, although it is less capable of satellite imagery and the census needs more effort.

## 8  An algorithm for recognition of palm shape

After the bounding of every palm as in the previous procedure, it is needed to recognize the palm shape:

1. Reading the separated shape of the single palm, figure (10, a). The green band is proper to this algorithm because of the predominant green color of the palm. Then by applying threshold filter to convert it to a binary image, and fill the gaps and neglect the remained as in figure (10, b).

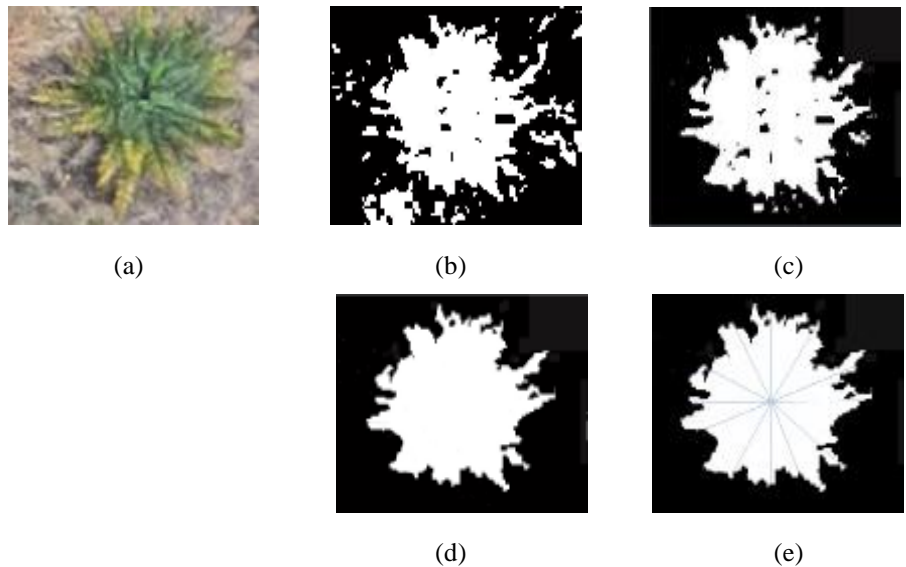

**Figure (10) Applying the algorithm to recognize the palm shape.**

2. The small pieces are considered as noise and deleted by a mean filter. The greatest part is the shape of the palm as in figure (10, c).
3. Appling an erosion filter to remove the inner black parts as in figure (10, d).
4. Measuring the geometric dimensions of the shape of the palm, where we choose the dimensions from the center of the image to the palm and at a distance of 30 degrees between the distance and another. These dimensions are geometric characteristics to determine the center of the mass and rotating from zero to 360 degrees, and measure the distance using Euclid's law to measure the distance between two points between the center and the boundary of the shape, until we recognize the whole shape of the palm, as in figure (10, e).
These steps are illustrated with the following figure:

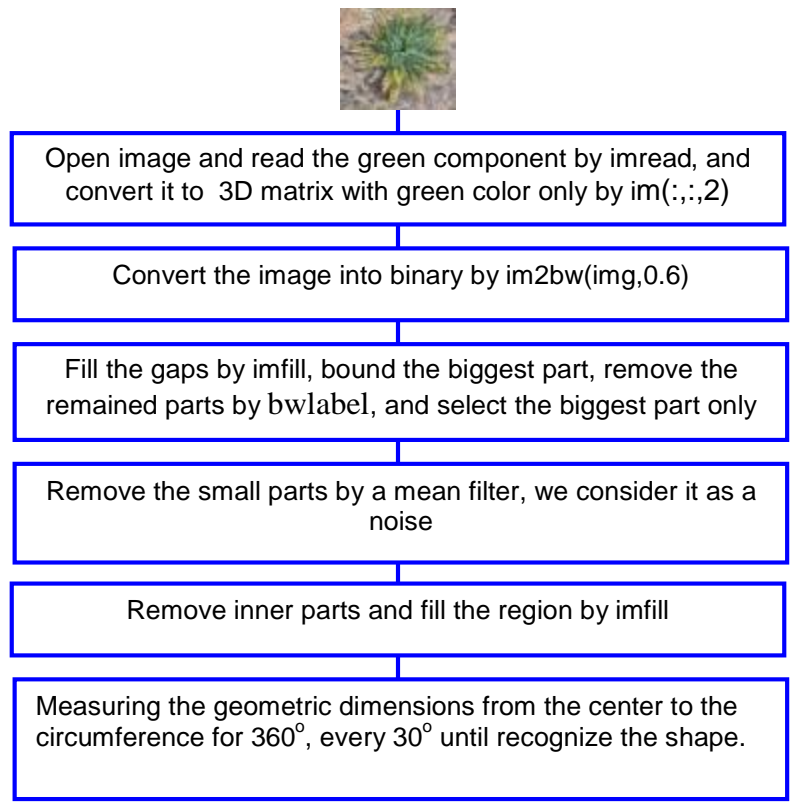

**Figure (11) A diagram illustrate the steps of recognizing palm shape.**

The recognizing of irregular small shapes is very difficult, so some researcher made it by manual steps with the image enhancement software [R. 8].

## 9     The Results and Allegations

We claim about this method used here:

1 - It is a mechanical method of calculating the number of palm trees from the satellite image programmatically.

2 - It provides the effort to count palm trees manually, and alternative to statistics from the ground.

3 - It is a new way to take advantage of the space image and distinguish the features in the picture of its form, where no one outside of Iraq because of the availability of palms in Iraq in large numbers are not in Iraq to be dominated by research, and I did not find my predecessors of the Iraqi researchers perhaps because of There is no need to provide statistical data in Iraq on the number of dates and the adoption of simple statistics.

4. It prepares to other researches to distinguish other forms of plants and others if the conditions we mentioned for palm trees are met, which do not overlap with each other and provide them with large internal shades that can be distinguished from the high-resolution image (0.6 m).

5. With more accurate satellite imagery, the same idea can be used to distinguish other features or trees, even large animals such as elephants, hippos and giraffes.

6 - This method can be applied to aerial images because they are of course more accurate, but the images of aircraft less capacity (R. 9) The cost of computation of the program in terms of effort and time is not very useful.

7- It can be applied to images taken from Google Earth explorer after downloading from Google [R. 10].

## 10    Conclusion

The algorithm designed in this research can be used to calculate the number of palm trees in QuickBird satellite images, which have (0.6 meters) resolution. For Other satellite image or different resolutions. There are trees of symmetric top view like pine trees, we can change the algorithm to proper the characteristic of these trees. With the developments of satellite sensors the distinguishing of self-shadow will be easier and more accurate, so this will enables to recognize the self-shadow of the shaded objects.

# References

[1]     Beginnings and early history of date palm garden cultivation in the Middle East, Tengberg, M., Journal of Arid Environments. Vol. 86, 2011, pps. 139-147.

[2]     Report of Food and Agriculture Organization (FAO) of the United Nations for 2013 & 2014, 23/2/2016. http://www.fao.org/faostat/en/#data/QC/visualize.

[3]     3- النخلة سيدة الشجر، عبد القادر باش أعيان العباسي، دار البصرى، ط1، ص 37.

[4]     Detecting Shadows In Quickbird Satellite Images, V. Arévalo , J. González a, J. Valdes b, and G. Ambrosio, 2005, p. 2.

[5]     Remote Sensing Sustainable Forest Management, Steven E. Franklin, Lewis Publisher, 2001, p. 20.

[6]     The Date Palm In Southern Nevada", M. L. Robinson, Associate Professor & Area Specialist, University of Nevada, pp. 2-3.

[7]     A Guide to MATLAB for Beginners and Experienced users", Brian R. Hunt. Cambridge University Press, 2$^{nd}$ Edition, 2006.

[8]     Classification of the Earth Cover for the Regions around the College of Science, I. A. Al-Helaly, College of Education Journal, in Thi Qar University, Vol. 2. Issue 1, 2012, pp. 278-295.

[9]     Remote Sensing and Image Interpretation, Thomas M. Lillesand, Wiley, 5$^{th}$ Edition, 2004.

[10]    Download the Google Earth from the web site: https://www.google.com/earth/.